\documentclass[12pt]{article} 
\usepackage[tablesfirst,notablist,nomarkers]{endfloat} 

\usepackage{OL}

\begin{document}
%
%

\title{Ultraslow optical waveguiding in an atomic Bose-Einstein condensate}
\author{Devrim Tarhan}
\affiliation{Department of Physics, Istanbul Technical University, Maslak 34469, Istanbul, Turkey \\
Department of Physics,
Harran University, Osmanbey Yerle\c{s}kesi, \c{S}anl\i{}urfa,
Turkey}
\author{Nazmi Postacioglu}
\affiliation{Department of Physics, Istanbul Technical University, Maslak 34469, Istanbul, Turkey}
\author{\"{O}zg\"{u}r E. M\"{u}stecapl\i{}o\~{g}lu}
\affiliation{Department of Physics, Ko\c{c} University, Rumelifeneri yolu,
Sar\i{}yer, Istanbul, 34450, Turkey}
\begin{abstract}
We investigate waveguiding of ultraslow light pulses in an atomic Bose-Einstein
condensate. We show
that under the conditions of off-resonant electromagnetically
induced transparency, waveguiding with a few ultraslow modes
can be realized. The number of modes that can be supported by the condensate
can be controlled by means of experimentally accessible parameters. 
Propagation constants and the mode conditions are determined
analytically using a WKB analysis. Mode profiles are found numerically.
\end{abstract}


An optical pulse can propagate at ultraslow speeds through 
an atomic Bose-Einstein condensate (BEC) \cite{hau} under
electromagnetically induced transparency (EIT) \cite{harris} conditions. 
Such pulses can be utilized for
storage of coherent optical information\cite{liu,phillips}.  Most of the early theoretical 
investigations assume one-dimensional light propagation\cite{morigi}, except for a 
few notable studies\cite{dutton,duan,ozgur}. 
Confinement of optical pulses in all three
directions has been examined quite recently\cite{andre}.
Transverse intensity profile of a
focused control pulse translates to an effective, spatially varying refractive index for the 
probe pulse, which can be used for waveguiding and confinement purposes\cite{andre}. Controlled spatial 
variation of the refractive index profile was first suggested and used for creation of
quantum shock waves with ultra-compressed slow light pulses \cite{dutton}. The shock wave induced snake instability, depending on the transverse density variations of the BEC, was observed \cite{dutton}. 
More recently, transverse effects on slow light 
have been studied numerically\cite{cheng}. It has been found that, using waveguiding by 
control pulses with Gaussian shaped transverse profiles, smaller group speeds can be achieved for higher order 
modes of the probe pulse\cite{cheng}. 

Possibility of slow light waveguiding in ultracold 
atomic medium has been
demonstrated in a recent experiment\cite{veng}. Decrease of refractive index away from the optical 
axis and the corresponding graded index optical waveguiding are 
consequences of Gaussian density profile of the ultracold atomic cloud, strongly confined
in transverse dimensions\cite{andre}. The experiment uses recoil induced resonance (RIR) method to 
achieve slow light ($c\sim 1500$ m/s), due to the strong dispersion in the high gain 
regime\cite{andre}. The ultracold atomic medium with RIR
scheme operates as a graded index optical waveguide 
in the strongly guided regime due to the large core radius ($\sim 200\,\mu$m) and high
refractive index contrast\cite{veng}. 

In this letter, we investigate ultraslow waveguiding by BEC under EIT conditions.
By considering slightly off-resonant EIT pulses propagating in a BEC, tightly trapped in 
transverse dimensions, possibility of weakly guided regime with few modes is shown. 
Single mode condition is established. These regimes are promising for guided nonlinear 
optical phenomena in ultracold
matter\cite{veng}. Multiple modes at low temperatures are found. These results
might be useful to design spatially controllable 
higher capacity optical memories \cite{cheng}.

%
At low temperatures a Bose gas can be considered as a condensed cloud in a thermal gas background. 
Atomic number density profile of such a system
can be described by\cite{naraschewski} $\rho(\vec{r})=\rho_c(\vec{r})
+\rho_{{\rm th}}(\vec{r})$, where $\rho_{{\rm c}}(\vec{r})=
[(\mu-V(\vec{r}))/U_0] \Theta(\mu-V(\vec{r}))$ is the number density of the condensed atoms
and $\rho_{{\rm th}}$ is the density of the noncondensed ideal Bose gas.
Here $U_0=4\pi\hbar^2 a_{s}/m$; $m$ is atomic mass; $a_s$ is the
atomic s-wave scattering length. 
$\Theta(.)$ is the Heaviside step function and $T_C$ is the critical temperature. The external
trapping potential is $V(\vec{r})=(m/2) (\omega_r^2 r^2+\omega_z^2
z^2)$ with $\omega_{r},\omega_{z}$ are trap frequencies for the radial and axial directions, respectively.
At temperatures below $T_c$, $\mu$, the chemical potential is determined by 
$\mu(T)=\mu_{TF}(N_0/N)^{2/5}$, where $\mu_{TF}$ is the chemical
potential evaluated under Thomas-Fermi approximation, 
$\mu_{TF}=((\hbar\omega_t)/2)(15Na_s/a_h)^{2/5}$, with 
$\omega_t = (\omega_z\omega_r^2)^{1/3}$  and $a_h=\sqrt{\hbar/(\omega_z\omega_r^2)^{1/3}}$, 
the average harmonic oscillator length scale. 
The condensate fraction is given by
$N_0/N=1-x^3-s(\zeta(2)/\zeta(3))x^2(1-x^3)^{2/5}$, with
$x=T/T_c$, and $\zeta$ is the Riemann-Zeta function. The scaling
parameter $s$ is given by
$s=\mu_{TF}/k_BT_C=(1/2)\zeta(3)^{1/3}
(15N^{1/6}a_s/a_h)^{2/5}$. 

EIT susceptibility\cite{harris} for BEC of atomic
density $\rho$ can be expressed as $\chi=\rho\chi_1$
with
\begin {equation}
\label{chieit} \chi_1 = \frac{|\mu|^2}{\epsilon_0 \hbar}
\frac{{\rm i}(-{\rm i} \Delta + \Gamma_2/2)}{[(\Gamma_2/2 
-{\rm i}\Delta)(\Gamma_3/2 -{\rm i} \Delta) + \Omega_c^2/4]},
\end {equation}
where $\Delta=\omega-\omega_0$ is the detuning of
the probe field frequency $\omega$ from the atomic resonance $\omega_0$. 
$\Omega_c$ is the
Rabi frequency of the control field; $\mu$ is the dipole matrix
element for the probe transition. $\Gamma_2$ and $\Gamma_3$ denote the dephasing rates
of the atomic coherence.
We consider a BEC of $^{23}$Na atoms with parameters \cite{hau}, $N=8.3\times10^6$, $\omega_{r}=2\pi\times69$ Hz, $\omega_{z}=2\pi\times21$ Hz, $\Gamma_3=0.5\gamma$, $\gamma=2\pi\times 10.01$ MHz, and $\Gamma_2=2\pi\times 10^3$ Hz.  
We take $\Omega_c=2.5\gamma$ and $\Delta=-0.1\gamma$
so that 
$\chi^{\prime}=0.04$ and $\chi^{\prime\prime}=0.0006$
at $T=42$ nK.
Here,
$\chi^{\prime}$ and $\chi^{\prime\prime}$ are the real
and imaginary parts of $\chi$, respectively. 
Neglecting $\chi^{\prime\prime}$, the refractive index becomes
$n=\sqrt{1+\chi^{\prime}}$. In the thermal cloud, $\chi^{\prime} \ll 1$ so that
an approximate refractive index profile in the radial direction can be written as
\begin{eqnarray}
n(r) = \cases{n_1[1-A(\frac{r}{R})^2]^{1/2} & $r\leq
R$. \cr 1 & $r\geq R$},
\end{eqnarray}
where
$A=1-1/n_1^2$ 
and $R=\sqrt{2\mu(T)/m\omega_r^2}$. In general $n_1$ is the index along the center line ($z$ axis) 
of the condensate. $n_1$ is different than $1$ only in the vicinity of the center of the condensate, 
where it varies slowly over the order of the optical wavelength. 
It can be determined by $n_1=(1+\mu\chi_1^{\prime}/U_0)^{1/2}$.
In the radial direction, BEC density profile is translated to a refractive index
analogous to a graded index fiber. Thermal
gas background plays the role of the fiber coating while the condensate acts as the core. 

We use the cylindrical coordinates as the refractive
index $n(r)$ is axially symmetric.
Small index difference between the condensed and noncondensed clouds results in weak guiding of
probe pulse, for which the normal modes are the linearly polarized modes (LP modes) \cite{gloge} which  
are determined by the Helmholtz radial equation 
\begin{eqnarray}
\label{diffequa}
\left[\frac{{\rm d}^2}{{\rm d}r^2}+\frac{1}{r}\frac{{\rm d}}{{\rm d}r}
+k_{0}^2 n^2(r)-\beta
^2-l^2/r^2\right ]\psi(r)=0.
\end{eqnarray}
Here $k_0=\omega/c,l=0,1,2,...$, $\beta$, and $\psi$ are defined for the transverse field of
the LP modes $E_t=\psi(r)\exp{[{\rm i}(l\phi+\omega t-\beta z))]}$.  
Employing Wentzel-Kramers-Brillouin (WKB) method, the 
characteristic equation is found to be
\begin{eqnarray}
\label{quant} 
\int_{r_{1}}^{r_2}[k_{0}^2 n^2(r)-\beta_{lm}
^2-\frac{l^2}{r^2}]^{1/2}{\rm d}r=(m+\frac{1}{2})\pi,
\end{eqnarray}
where $r_{1,2}$ are the zeros of the integrand and $m=0,1,2,3,...$. 
Solving Eq. \ref{quant} for $\beta$, we find
\begin{eqnarray}
\label{proconstant} \beta_{lm}=n_1
k_0[1-\frac{2\sqrt{A}}{n_1 k_0 R}K]^{1/2},
\end{eqnarray}
where $K=(l+2m+1)$. The propagation constant $\beta_{lm}$,
depends on the frequency of the probe pulse. This frequency dependence
determines the dispersion of the waveguide. 
Propagation constant can also be defined in terms of an effective mode index
$n_{lm}$ such that
$\beta_{lm}=n_{lm}k_0=n_{lm}(n_1,\omega)\omega/c$. 

\begin{figure}[htb]
\centerline{\includegraphics[width=8.3 cm]{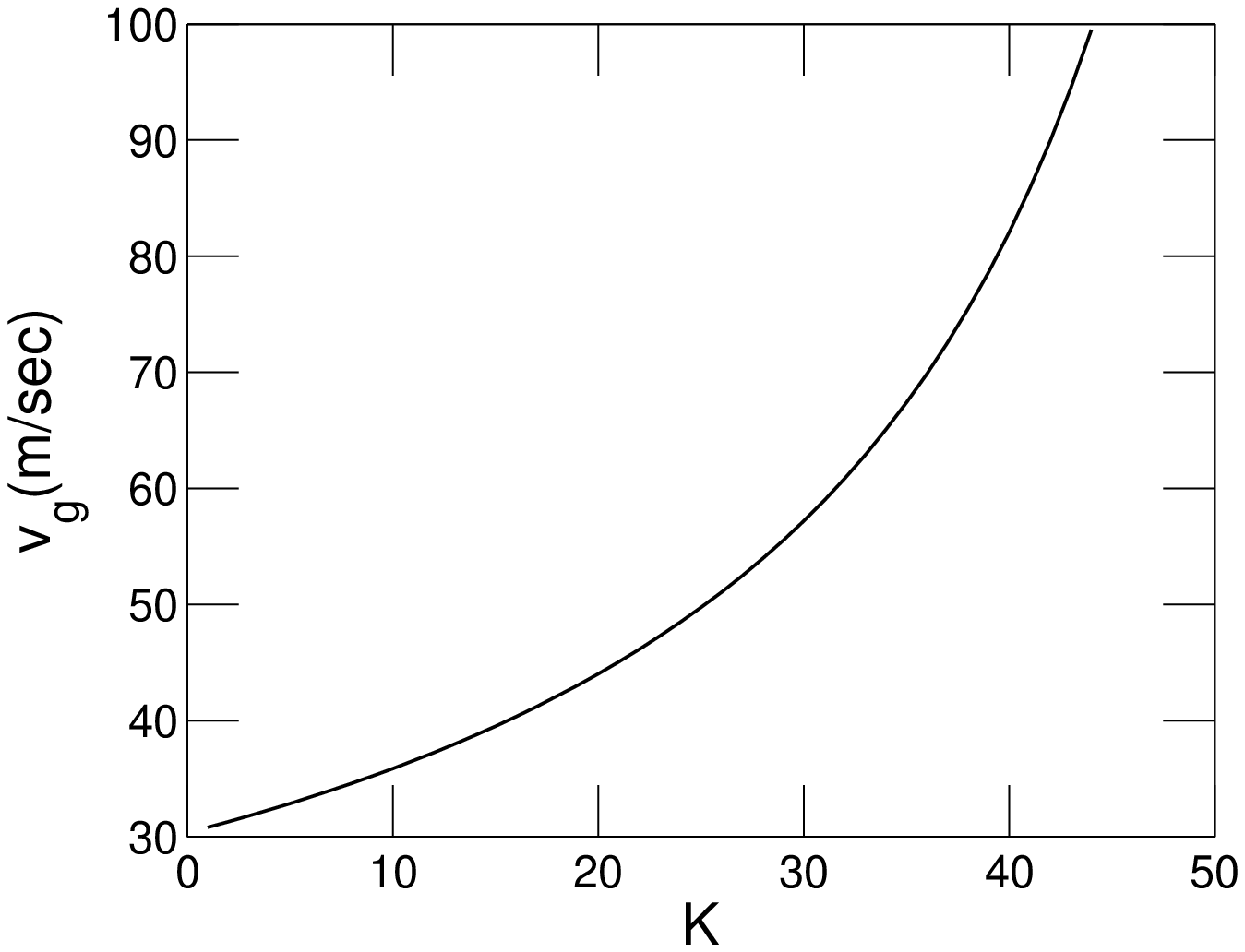}}
\caption{Dependence of the group velocity on $K=(l+2m+1)$ at
$T=42$ nK ($T_c=424$ nK). 
Parameters are given in the text. 
For the single mode condition group velocity is
$30.8$ m/s. } 
\label{fig1}
\end{figure}

The group
velocity $v_g$ is defined as $ 1/v_g ={\rm d}\beta/{\rm d}\omega$,
which leads to 
\begin{eqnarray}
\label{groupvelocity} \frac{1}{v_g} =\frac{d \beta}{d
\omega}=\frac{\omega}{c}\left ( \frac{\partial n_{lm}}{\partial
n_1}\frac{\partial n_1}{\partial \omega}
+\frac{\partial n_{lm}}{\partial \omega} \right )+\frac{n_{lm}}{c}.
\end{eqnarray} 
Here, the terms in the parenthesis are due to the
material and waveguide dispersion, respectively.
Under EIT conditions, material
dispersion dominates. Furhtermore, $n_{lm}\approx 1$ so
that we can write
\begin{eqnarray}
\label{groupvelocity2} v_g = \frac{c}{1+\omega \frac{\partial
n_{lm}}{\partial n_1}\frac{\partial n_1}{\partial \omega}}.
\end{eqnarray}
The group velocity at $T=42$ nK as a function of $K$ is given 
in Fig.\ref{fig1}. We calculate $v_g=30.8$ m/s,
for the LP$_{00}$ mode. Group velocity increases with the mode number $K$. 
The higher order modes are away from the high density region of the condensed
cloud and move faster. This behavior
reduces modal dispersion and can be beneficial to preserve shape of the 
probe pulse for coherent optical information applications in BECs.

\begin{figure}[htb]
\centerline{\includegraphics[width=8.3cm]{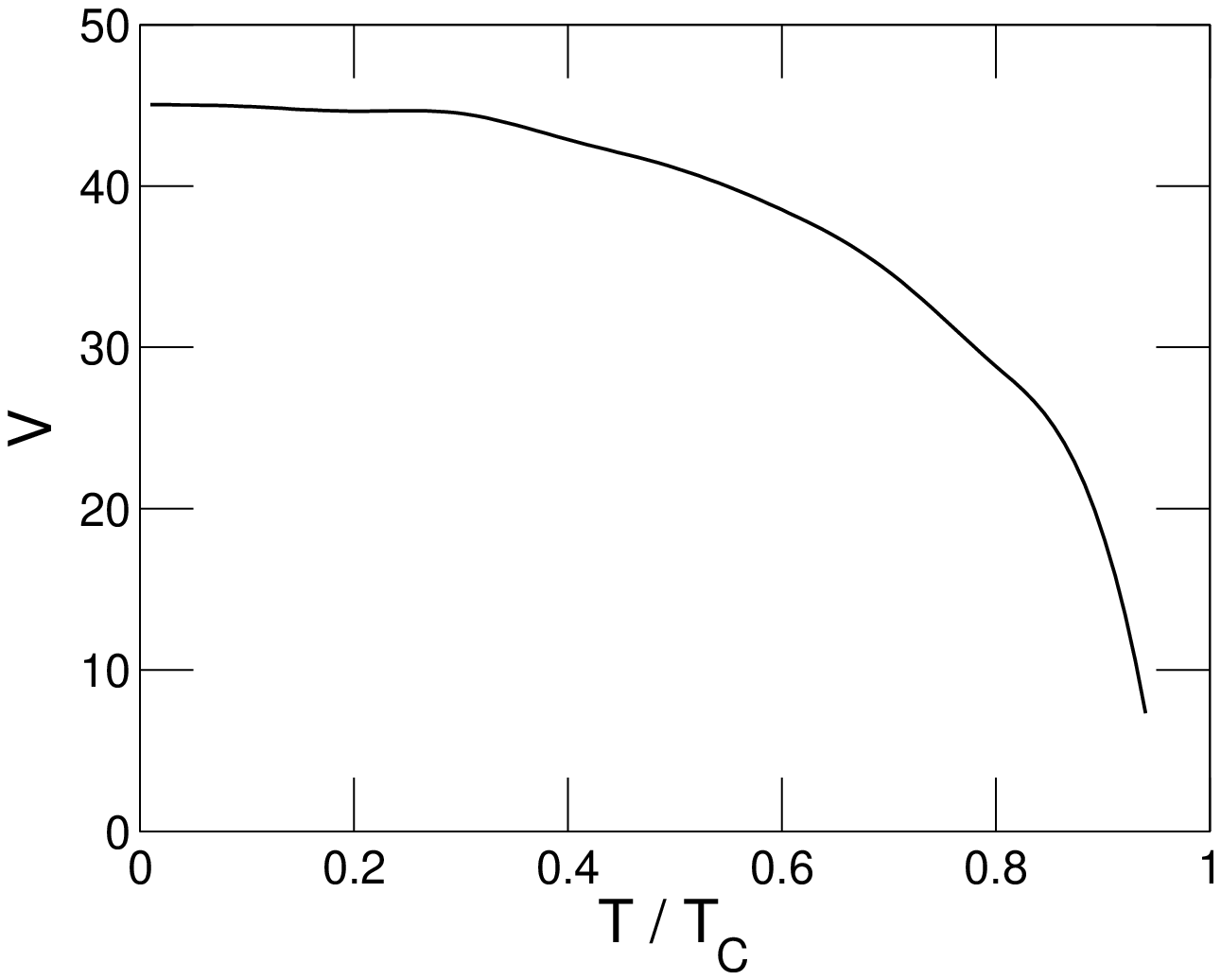}}
\caption{Temperature dependence of $V$ is plotted
from $T=42$ nK to $T=398.7$ nK ($T_c=424$ nK). $V$ is related to $K$ via $K\le V/2$.} 
\label{fig2}
\end{figure}

The dimensionless normalized frequency $V$
is defined as $V=n_1k_0R\sqrt{A}=k_0R(n_1^2-1)^{1/2}$.
It determines the number of modes that can be supported by the atomic cloud.  
We find a confined mode description via $k_0<\beta<n_1k_0$ so that
$K\leq  V/2$. 
A condensate supports only single LP$_{00}$ mode for
$2\leq V<4$. For $2\leq V<6$, it could support just two
modes, LP$_{00}$ and LP$_{10}$. $V$ can be tuned by various experimental  parameters as
$V=k_0\mu(2\chi_1^{\prime}/m\omega_r^2u_0)^{1/2}$. Multiple-mode support can be 
perhaps most conveniently tuned by $T$.
Temperature dependence of $V$ is plotted in
Fig.\ref{fig2}. For the parameters used in Fig. \ref{fig1}, at $T=42$ nK
it starts from $V=45$ and decreases down to 
a lowest value of $V\approx 7.3$ at $T=398.7$ nK. As temperature increases,
the condensed cloud shrinks and the index contrast diminish
so that the BEC can no longer support multiple mode propagation. 
%

\begin{figure}[htb]
\centerline{\includegraphics[width=8.3cm]{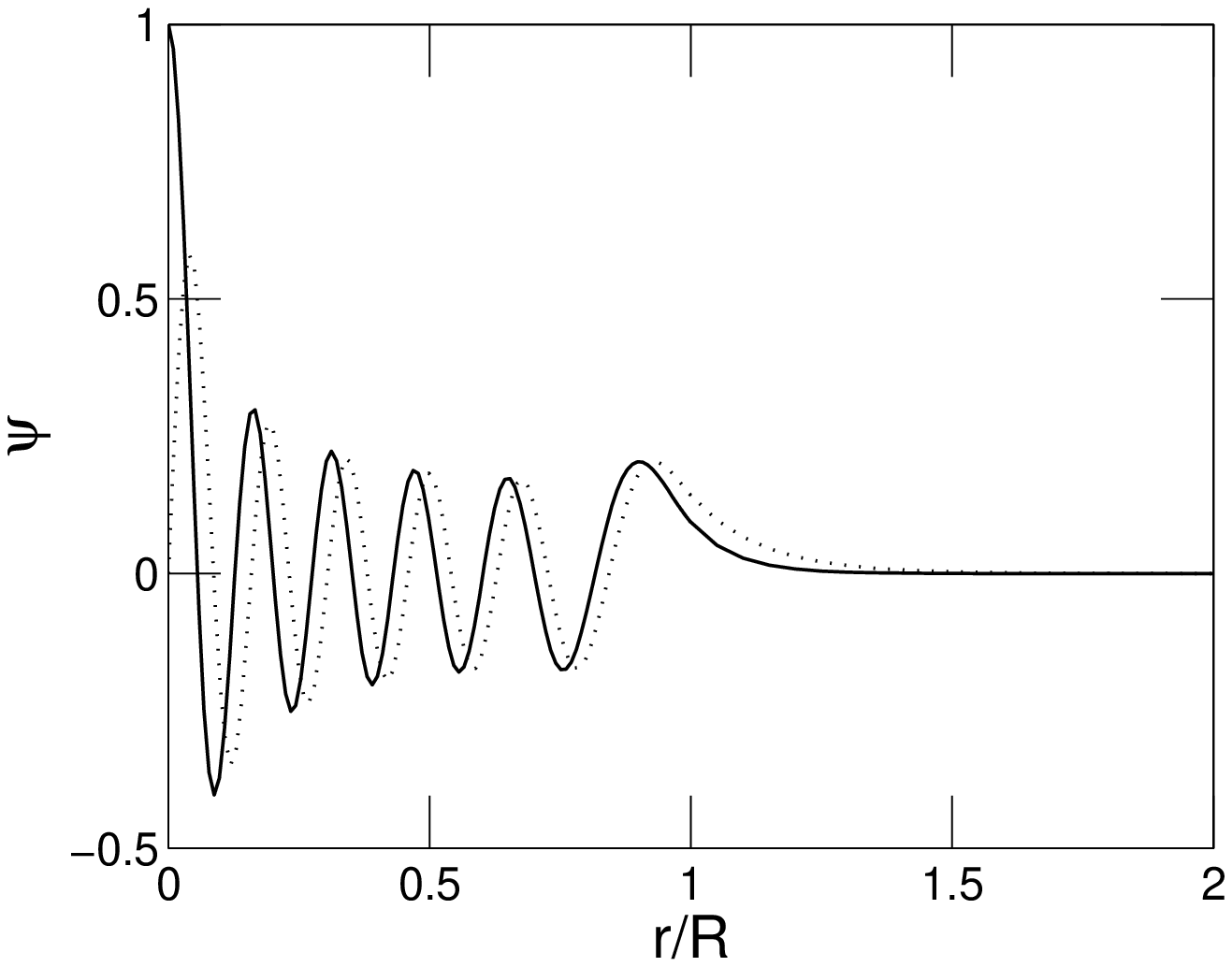}}
\caption{Radial profiles of transverse modes LP$_{00}$ (solid line) and LP$_{10}$ (dotted line)
confined in a BEC of $^{23}$Na atoms with parameters given in the text.} 
\label{fig3}
\end{figure}

More accurate results may be obtained using a numerical scheme.
For that purpose it is assumed that the index of refraction
increases incrementally from the edge to the center of the core.
If the number of increments is high enough, this model will be a
faithful representation of the conditions prevailing in the
inhomogeneous core. Within each increment the refractive index
is constant and we can use exact solutions of the wave equation in terms of 
Bessel functions. Electromagnetic boundary conditions 
provide a recurrence relation between successive solutions as well as
an equation for the propagation constant. We have compared
the results of our numerical calculation with these of WKB
calculations and found excellent agreement. Typical results
of our numerical simulations are given in Fig.\ref{fig3}, where
the modes LP$_{00}$ and LP$_{10}$ are shown. The modes are localized in 
the condensate core of radius $R=21.1\,\mu$m. Beyond $R$, in the thermal component, 
the modes are evanescent. Such modes can be addressed by specifically constructed transverse 
profiles of the probe pulses. This can be exploited to control of capacity in the 
transverse direction for storage of coherent optical information in different mode patterns. 
These results can be further combined with the longitudinal control of optical information 
storage capacity where a particular EIT scheme, in which the control field is perpendicular 
to the probe field, is employed \cite{dutton}.    
%

Summarizing, we have examined optical waveguiding of ultraslow pulses in a Bose-Einstein
condensate under EIT conditions. Off-resonant EIT scheme has been
considered to permit multi-mode ultraslow light propagation. Propagation
constant is calculated analytically using WKB approximation. Mode profiles 
are determined numerically.
Single and two mode conditions are established in terms of experimentally
accessible parameters. Temperature dependence of the number of modes that can be
supported by BEC is presented. Such ultraslow modes may be useful for realizing
spatially controllable storage of coherent optical information with higher capacity
in BECs, in the transverse direction, complementing longitudinal control of information storage capacity \cite{dutton}. 

\"O.E.M. acknowledges support by T\"UBA-GEB\.{I}P Award.  D.T.
is supported by Istanbul Technical University Foundation (ITU BAP).

\end{document}